\documentclass[conference]{IEEEtran}

\usepackage{cite}
\usepackage{amsmath,amssymb,amsfonts}
\usepackage{array}
\usepackage{booktabs}
\usepackage{graphicx}
\usepackage{url}
\usepackage{xcolor}
\usepackage[hidelinks]{hyperref}
\usepackage{orcidlink}
\usepackage{seqsplit}
\newcommand{\hex}[1]{\texttt{\seqsplit{#1}}}

\begin{document}

\title{panvram: GPU-resident random access to a human pangenome cohort}

\author{%
\IEEEauthorblockN{Yakiv Shavidze~\orcidlink{0009-0008-3622-3448}}
\IEEEauthorblockA{Independent researcher\\
ORCID: 0009-0008-3622-3448\\
aceapex.contour@outlook.com}%
}

\maketitle

\begin{abstract}
Repeatedly sampling short sequence windows from a human pangenome requires an
access representation as well as compression. panvram stores each assembly as a
reference-relative block archive and pools these archives on a GPU. For 558
haplotype assemblies, the default q4k representation occupies 16.642~GB of
persistent pools including the decoded reference; a packed-reference,
compact-directory representation occupies 11.634~GB. These quantities are not
peak application memory. In the C1 experiment, default and fully compact forms
delivered 743{,}271 and 736{,}762 windows/s; a separate public Colab run
delivered 741{,}931 and 680{,}760 windows/s. All are batch API measurements and
must not be pooled as repeated trials of an identical environment. Full-decoding
checks compared reconstructed FASTA checksums with the manifest. On the CPU, the
M6 shuffled-window workload at the same cohort cardinality reported 26
requests/s for AGC and 145{,}222 requests/s for refrel3 q4k. This comparison
concerns one-thread requests of 4~KiB on the recorded warm-cache workload, not
all genome-access patterns. AGC provides the smaller incremental archive:
5.40~MB/assembly versus 13.04~MB/assembly for refrel3 q16k, with different
exclusions; the size pair is not the same refrel3 configuration as the q4k
timing pair. We derive an expected-work expression for fixed independent
granules and distinguish it from an empirical median-latency predictor.
Cohort-scale counter and DRAM measurements motivate a locality hypothesis, but
do not establish a causal per-miss latency or independently reproduce a
model-failure verdict. The window-cost verdict at $N=558$ is published with its
harness and raw timings.
\end{abstract}


\section{Scope, data and contribution}

The principal object is a manifest-defined cohort of 558 haplotype assemblies
encoded against a shared T2T reference~\cite{ref:t2t}. $N$ counts assemblies, not
people. HPRC~\cite{ref:hprc} is identified through BioProject
PRJNA730823;\footnote{%
\url{https://www.ncbi.nlm.nih.gov/bioproject/PRJNA730823}}
the BioProject is an umbrella resource, not the manifest of this experiment. Use
of its data does not imply HPRC endorsement, participation in this benchmark, or
approval of the manuscript. Source accessions, releases and checksums must be
retained per manifest row.

The contributions are GPU-resident cohort access, an explicit accounting of
representation costs, and a comparison of access workloads with precisely scoped
correctness checks. The mathematical contribution is an expected-granule
calculation and a falsifiable calibrated predictor, a cost model in the sense
of the roofline analysis~\cite{ref:roofline}, not a general statement about
latency. The historical $N=4$ experiment is retained in Appendix~A without
changing its PASS/FAIL decisions. It is not a main experimental result or a new
small input dataset.

This manuscript measures a single host and a single cohort. Number provenance,
source hashes and unresolved claims are recorded in the evidence package
(Sec.~IX). The $N=558$ window-cost verdict was re-measured on 2026-10-17 with a
public harness; no other benchmark was rerun. The smaller-cohort predictor
experiment is explicitly secondary.

A companion measurement of block granularity across four compressed formats on
one host is reported separately~\cite{ref:granularity}. The present manuscript
adds the GPU-resident cohort representation, the cohort-scale ($N=558$) locality
effect that the sequential rate does not capture, and the AGC comparison. The
window model and its model-B acceptance rule are shared between the two.

\section{Representation, coordinates and correctness}

\subsection{Independent blocks and dependencies}

refrel3 v1 stores each assembly separately as copies against the decoded
reference plus entropy-coded literals, as specified in the repository
\texttt{FORMAT.md}. Block decoding depends
on the shared reference, archive metadata and the selected block, not on a
preceding decoded block. Lower-case runs and FASTA framing are reconstructed
separately. The metadata includes contig geometry and block state; the header
binds the reference by length and SHA-256. The q4k and q16k configurations use
canonical granules of 4{,}096 and 16{,}384 bytes. These are sequence bytes, not
compressed bytes or serialized FASTA bytes. A decoder must reject invalid
section bounds, unsupported flags, invalid copy coordinates and malformed entropy
streams.

Requests use the assembly's own contig and zero-based half-open coordinates
$[\text{start},\text{end})$. They are not queries in reference coordinates and do
not provide a coordinate-liftover service. Upper-case canonical sequence bytes,
case-preserving window bytes and complete FASTA files are distinct output
domains; the judge must specify which domain it compares.

\subsection{Integrity checks, window checks and corruption scope}

SHA-256 binds source files and benchmark answers to the expected inputs. XXH3 is
a non-cryptographic consistency checksum: when block checks are present and
enabled, the CPU verified path checks decoded blocks; full decoding can also
compare the reconstructed FASTA checksum with the header and manifest. A regional
request cannot verify a full-FASTA checksum without reconstructing that FASTA.
The GPU path's lack of block-XXH3 checking must not be hidden behind CPU
integrity results.

Window correctness is checked separately against the source FASTA or an
independently extracted truth window, using the specified coordinates and output
domain. CPU/GPU equality compares implementations; source-FASTA equality checks
against the input. Neither should be described as the other. A coordinate-shift
negative control must fail where the expected bytes differ.

The historical corruption harness attempts a full CPU decode, not a regional
fetch. Its external oracle is the checksum of reconstructed FASTA; a harmless
classification means agreement under that checksum oracle, not automatically a
separate byte-by-byte comparison. Changing both hash presence and verification
mode does not isolate the effect of block hashes. This version describes the
verification mechanism only and does not retain numerical mutation-campaign
outcomes as headline evidence.

To reproduce that campaign, its archive hash, seed, exact command, process
count, timeout and mutation distributions must accompany the record. The legacy
category named \emph{payload} selects offsets after the first 1{,}024 file bytes,
not necessarily within a parsed payload section. It should be described as
post-prefix damage unless section membership is actually enforced. No general
detection guarantee, regional detection rate or GPU corruption result is
inferred.

\subsection{Size accounting}

MB and GB are decimal units; KiB, MiB and GiB are binary units. One canonical
byte stores one upper-case base; canonical byte counts and base counts are
therefore the same number, and a rate in Gbp/s equals the same rate in canonical
GB/s. $W$ and $Q$ are counts of canonical bytes; windows quoted in bases (8{,}192;
65{,}536; 4{,}096) are the same counts. An incremental size is explicitly a
subtraction followed by division by cohort size, not the independently usable
size of one assembly archive.

For the cohort of 558, size accounts are as in Table~\ref{tab:sizes}. Complete
refrel3 archive bytes still exclude the required external reference; AGC's
complete archive includes T2T. The incremental columns therefore do not
constitute a matched total-deployment footprint.

\begin{table}[t]
\caption{Cohort size accounts ($N=558$). Incremental MB/assembly is a
subtraction over cohort size, not a standalone archive size; exclusions differ
by row.}
\label{tab:sizes}
\centering
\footnotesize
\setlength{\tabcolsep}{3pt}
\begin{tabular}{@{}lrrr@{}}
\toprule
Configuration & Complete & Excluded & Incr.\ MB/ \\
 & archive bytes & bytes & assembly \\
\midrule
refrel3 q4k  & 12{,}330{,}370{,}123 & 3{,}282{,}107{,}128 & 16.22 \\
refrel3 q16k & 8{,}097{,}986{,}851  & 820{,}528{,}488    & 13.04 \\
AGC with T2T & 3{,}721{,}635{,}360  & 707{,}323{,}301    & 5.40  \\
\bottomrule
\end{tabular}
\\[2pt]
\raggedright\scriptsize
Complete MB/assembly, rounded: refrel3 q4k 22.10, refrel3 q16k 14.51, AGC 6.67.
Excluded bytes are block hashes for refrel3 and the separate T2T-only archive
for AGC.
\end{table}

The deployment account is completed by the reference distribution file used by
refrel3 and by the GPU pools. The decoded T2T reference \texttt{t2t.fa} is
3{,}156{,}259{,}565 bytes, SHA-256
\hex{51ab7d3c2a375ecff53ee2a1be03e58d57cc1601c948a860c03a78d724f26e4c}; the
decoded canonical reference inside the archives is 3{,}117{,}292{,}070 bases,
bound in every archive header. refrel3 needs no index or sidecar file: the index
is built in memory when encoding. Table~\ref{tab:sizes} is not a claim of matched
verification settings.

\section{GPU-resident cohort results}

\subsection{Persistent pools, not peak VRAM}

The default q4k pool has 16{,}642{,}219{,}335 bytes: reference
3{,}117{,}292{,}070; payload 8{,}448{,}112{,}345; block table
4{,}923{,}167{,}388; model tables 153{,}647{,}532. The byte-level sum is exact;
its rounded decimal display is $16.642 = 3.117 + 8.448 + 4.923 + 0.154$~GB. These
CPU-path pool accounts are complemented by the GPU measurements below.

Packing the reference and compacting the block table gives
11{,}634{,}041{,}102 persistent bytes, or 11.634~GB rounded.
\texttt{resident\_bytes()} measures persistent pools after open. Output tensors,
part of the scratch allocation, framework/runtime overhead and model-training
memory are outside that account. The compact experiment ran on an
A100-SXM4-80GB, not on a physical 16~GB GPU; no end-to-end deployment on a
smaller card is claimed.

\subsection{Separate sampling experiments}

Both reported sampling experiments use batches of 1{,}024 windows of 8{,}192
bases and report the median of five calls. They are distinct runs. A rate is
windows divided by elapsed batch-API time, not the inverse of the p50 latency of
individual windows (Table~\ref{tab:sampling}).

\begin{table}[t]
\caption{Separate sampling experiments. C1 and public Colab are distinct runs
and are not pooled.}
\label{tab:sampling}
\centering
\footnotesize
\begin{tabular}{@{}llrr@{}}
\toprule
Experiment & Form & Persistent GB & Windows/s \\
\midrule
C1            & default          & 16.642 & 743{,}271 \\
C1            & packed reference & 14.304 & 674{,}822 \\
C1            & compact block table & 13.972 & 710{,}235 \\
C1            & both compact     & 11.634 & 736{,}762 \\
Public Colab  & default          & 16.642 & 741{,}931 \\
Public Colab  & both compact     & 11.634 & 680{,}760 \\
\bottomrule
\end{tabular}
\end{table}

C1 is the log dated 2026-10-07; public Colab is the separately retained notebook
output dated 2026-10-09. The public Colab file was copied from notebook output
by the author, not emitted by an automatic log writer. This provenance
distinction is retained.

The measured sample boundary is wall-clock around
\texttt{sample(..., return\_coords=True)}, bracketed by CUDA synchronisation. It
includes coordinate generation and API work; it is not a decode-kernel-only
timer. In the reviewed script, a warm-up precedes the timed loop, and the CPU
equality check occurs after the loop, checking only the last timed batch. It
does not certify every timed batch. Separate fetch and full-decode checks retain
their own scope.

The fetch test uses random variable-length intervals rather than the
fixed-window workload of the CPU predictor. The reviewed script uses seed
20261005 for sampling and fetch generation, chooses assemblies uniformly and
contigs with probability proportional to length, then sets
$L=\min(\text{contig\_length},\,1+\lfloor 2^{20U}\rfloor)$ for uniform $U$ in
$[0,1)$. It chooses a valid start uniformly. Each fetch timer is bracketed by
CUDA synchronisation; host transfer for the equality check occurs afterwards.
These script facts must be bound to the exact executed revision, especially for
the separate public notebook. GPU batch throughput and CPU individual-window
quantiles must not be directly ranked as if they shared a timer and workload.

\subsection{Full decoding and end-to-end validation}

Each C1 form reports full decoding of 558 assemblies with FASTA XXH3 matching
the manifest, and 1{,}000 fetches matching the CPU decoder. These finite checks
are separate from the last-batch sampling comparison.

The public Colab default run reports a summed GPU-decode-plus-host-copy interval
of about 2.5 minutes and 11.05~Gbp/s, versus 10.1 minutes for the end-to-end
verified pass; its compact run reports about 2.5 minutes, 11.01~Gbp/s and 10.2
minutes end-to-end. Rebuilding FASTA and hashing are part of the longer path.
Calling the shorter interval pure GPU kernel time would be incorrect.

C1 instead reports default 2.5 minutes for decode-plus-copy and 10.5 minutes
end-to-end; the fully compact form reports 2.6 and 10.6 minutes. The default C1
rate is 11.02~Gbp/s, not the public Colab value. Rounded times must not be used
to reconstruct rates more precisely than the original log. No single combined row
mixes these experiments.

\section{Window cost model: derivation and limits}

\subsection{Geometry and units}

Let $Q$ and $W$ be positive integer counts of canonical bytes: $Q$ is the fixed
independently decoded granule length and $W$ the requested window length. Let $U$
be the starting residue, uniform on $\{0,\dots,Q-1\}$. Assume no contig-end
truncation, fixed-size blocks and complete decoding of every touched block. Let
$D_Q > 0$ be a sequential canonical-output rate in bytes/s and $c_0$ an additive
time in seconds. The constant $1$ in $W+Q-1$ denotes one canonical-byte position
in this discrete calculation.

The touched-block count is $K(U,W)=1+\lfloor (U+W-1)/Q\rfloor$. Write
$W-1=aQ+r$, with $0\le r<Q$. Then $K=a+1$ for $Q-r$ residues and $K=a+2$ for $r$
residues. Thus $E[K]=a+1+r/Q=1+(W-1)/Q$ and $E[QK]=W+Q-1$.

If, additionally, each request obeys $t(U,W)=c_0+QK(U,W)/D_Q$ with constant
per-byte cost and overhead, linearity of expectation gives exactly
\begin{equation}
E[t(W)]=c_0+\frac{W+Q-1}{D_Q}.
\end{equation}
Its unit is seconds: canonical bytes divided by canonical bytes/s plus seconds.
This is a conditional mean-time result, not a derivation of measured median
latency.

The distinction matters: the lower median of $K$ is $a+1$ when $2r\le Q$ and
$a+2$ otherwise. It generally differs from $E[K]$. Data-dependent decode work,
memory latency, block boundaries, scheduling and per-query overhead further
invalidate an automatic expectation-to-median substitution.

\subsection{Empirical predictor and unchanged acceptance rule}

The historical protocol tests the empirical prediction
$\hat{t}(W)=c_0+(W+Q-1)/D_Q$ against measured p50. With $m_1=\text{p50}(W=1)$ from
a separate calibration and $c_0=m_1-Q/D_Q$, substitution yields
$\hat{t}(W)=m_1+(W-1)/D_Q$. $Q$ cancels explicitly conditional on $m_1$ and
$D_Q$, although those measured quantities and the sign condition on $c_0$ may
depend on $Q$. The intercept is a calibrated residual, not an independently
measured physical overhead.

The historical verdict windows are 1{,}024, 8{,}192 and 65{,}536 canonical
bytes, with 10{,}000 requests at each window, seed 20261003 and one persistent
handle in one-thread execution. p50 is the nearest-rank median. The job uses
$\text{error}=100\times(\text{measured\_p50}-\text{prediction})/\text{prediction}$.
Model B requires absolute error $\le 20\%$ at every verdict window and
$c_0\ge 0$. The calibration window is not a verdict point. These thresholds and
historical PASS/FAIL decisions are unchanged.

The protocol v1.1 of the hardware window job is
\texttt{review/axis3/PROTOCOL\_AXIS3.md}, SHA-256
\hex{4a0027fec59b3252dc07b29458a57a5a451957f1b64b36f3927765cbc6be7a5c}. The
20\% tolerance, $c_0\ge 0$ and the verdict windows 1{,}024 / 8{,}192 / 65{,}536
are in that protocol and were fixed before the 2026-10-08 run. We state a
pre-registration caveat: the run performed the full sweep
$W=1,\,2,\,4,\,\dots,\,65{,}536$ (powers of two, 17 points) instead of the
minimal four-window list that the later default runbook describes; the choice of
the full sweep was not frozen in writing before the run (the runbook difference is
documented in \texttt{logs/verdict-n4/RUNBOOK\_DIFF.md}\footnote{%
\url{https://github.com/yasha1971-coder/panvram/blob/8f8e9e1ccc328f07bc3cbfe40f0317db33e101a4/logs/verdict-n4/RUNBOOK_DIFF.md}}).
The verdict uses only the three protocol windows; the
extra points are diagnostics. The separate \texttt{validate-inputs} and local
native checks were not run as separate steps (the job's own input checks and CI
ran them).

Variable block sizes and non-uniform start residues require their own weighting.
An unweighted mean granule size does not turn a variable-geometry reader into the
exact fixed-$Q$ model. Neither $Q/W$ nor negative $c_0$ alone establishes a
general applicability boundary or an upper bound; the quantity $W+Q-1$ is the
read amplification of a fixed-granule store~\cite{ref:rum}. Historical
BGZF~\cite{ref:bgzf} and refrel3 rows include passes when $Q$ exceeds $W$, whereas
seekable zstd~\cite{ref:zstd} fails at a window larger than $Q$.

\subsection{Timing boundaries and historical results}

$D_Q$ must declare both its numerator and timed operations. Native decoding
directly to canonical bytes and native reconstruction of serialized FASTA with
framing stripped outside the timer can share a canonical numerator while
performing different work inside the timer. Allocation, copying, formatting and
result materialisation must be audited per adapter. The existence of an operation
does not establish its share of elapsed time.

The historical smaller-cohort table is retained in Appendix~A, not promoted into
the main cohort result. No refitted intercept or new verdict is inferred from it.
Its diagnostics are not extra acceptance points. Model A exceeded measured medians at the listed
LZ4~\cite{ref:lz4}/OZSEG points; this is not a mathematical bound for other
inputs. The largest listed
absolute OZSEG model-B error is 12.1\%, not strictly within 12\%. Error
percentages use prediction as denominator, not measured time.

\subsection{Cohort-scale locality and the reported model miss}

The reported miss at the 64~KiB window for $N=558$ requires its own verdict
evidence. That verdict is now published. Window-cost verdict at $N=558$ by
\texttt{tools/window\_cost} (panvram 36870bd), a one-file harness on the public
v1 reader: $D_Q$ 3.544~Gbases/s, $c_0$ 5.936~us, model-B error $+48.98\%$ at
64~KiB (FAIL), $-5.98\%$ and $+2.91\%$ at 1 and 8~KiB (PASS). The run covered 558
assemblies / 1{,}680{,}437{,}743{,}110 canonical bases at $Q=16{,}384$ canonical
bytes (q16k), with 10{,}000 requests per window (seed 20261003); every answer's
SHA-256 equalled truth, with 0 bad. The verdict is FAIL under model~B
($c_0\ge 0$ and absolute error $\le 20\%$ at every verdict window).

Three scope statements accompany this number. (a) The harness is an in-process
C++ fetch loop, not the Python job of the $N=4$ verdict. (b) v1 block-hash checks
were off inside the timer; the truth table was produced with every block hash
checked. (c) Harness source, build command, requests, truth and raw timings are
public, so the verdict is independently recomputable from the published
archives.

The available mechanism experiment compares the same 10{,}000 windows of
65{,}536 bases in random order and sorted by assembly/contig/start, using q16k
and one thread pinned to CPU~3. It checks answers against truth. These counters
document an order effect; by themselves they do not reconstruct the predictor's
$c_0$, $D_Q$, p50 error and PASS/FAIL.

Per-window counters are obtained by subtracting the open-plus-one-pass run from
the open-plus-eleven-pass run and dividing by the ten additional passes and the
request count. Group g1 reports about 111{,}822 additional cycles and about 289
additional local-DRAM demand fills per window in random order, commonly rounded
to approximately 290. Other counter groups use different events and must not be
mixed into this ratio.

The reported ratio is approximately 387 additional cycles per additional demand
fill, or 73.5~ns at the representative 5.26~GHz. This is an incremental
cycles/event ratio between workloads, not a direct per-access DRAM latency.
Concurrent misses, overlap, page translation, prefetching and other changed work
can all affect it. Multiplying a count by a separate pointer-chasing latency does
not establish the causal contribution to a latency median.

The separate pointer-chasing experiment on EPYC 4344P uses random cyclic
permutations of cache-line-sized nodes, CPU~0, three warm-ups and nine
measurements per working set. The median is 103.290157~ns/access at 256~MiB and
115.785740~ns/access at 4~GiB, conventionally summarised as about
103--116~ns over that large-working-set range. This is not the value for every
size: the recorded sweep starts at 32~KiB.

The DRAM results table was produced by \texttt{run.sh} (SHA-256
\texttt{b805687d\dots}), published in panvram
\texttt{logs/dram-2026-10-15/}; the wrapper provenance is reconciled, and the
harness is reproducible from that record.

\section{CPU comparison with AGC}

\subsection{Matched workload and limits}

M6 uses AGC~\cite{ref:agc} 3.2.4 at e67e3fc and refrel3 v1, on AMD EPYC 4344P with
the recorded 8-core/16-thread topology, Ubuntu 22.04.5 and g++ 11.4.0. The AGC build uses
\texttt{-O3 -march=native}. Archive creation uses
\texttt{agc create -d -k 31 -l 20 -s 60000 -b 50 -t 16 -o agcN.agc <T2T> <N
FIFOs>}, with the FIFOs fed from the q16k v1 archives in manifest order and T2T
first. The single-thread timing rows use one handle per thread, prefetching
$=1$ and affinity \texttt{taskset -c 3}; libagc is at e67e3fc. Every counted
request takes the \texttt{decompress\_contig} path with \texttt{fast = false}
(\texttt{dc\_fast\_false == dc\_calls}). Archive open is separate from the request
timer. Storage hardware remains unspecified. The M6 refrel3 and mech558 harness
is the public \texttt{tools/rr3bench} (panvram 8f8e9e1); a rebuild from the
published sources matches the logged binary byte-for-byte (SHA-256
\texttt{f2f84fe9\dots}), so these measurements are independently recomputable
from the published request generator (\texttt{make\_req.py}, seed) and input
hashes.

The warm-cache policy is repeated access after three warm-up passes, followed by
nine timed runs; it does not imply the whole cohort fits in CPU caches. Each
window run contains 10{,}000 requests of 4{,}096 bases in shuffled sample order,
and the logs report every answer checked by SHA-256. For the principal cohort,
the aggregate one-thread rates are 26 requests/s for AGC, 145{,}222 for refrel3
q4k and 73{,}690 for q16k. Thus the 13.04~MB q16k size must not be paired with
the q4k speed as if both described one configuration.

The request seeds depend on $N$: \texttt{random.Random(20261011+N)} for windows.
Increasing $N$ changes the archive, available sequences and actual request
coordinates together. The series is not a controlled experiment varying only
cohort size. Counters summarise the first 1{,}000 requests, whereas timing covers
the full window series; their medians need not represent the same individual
requests.

\subsection{Timing versus work counters}

For each timed run, completed requests divided by elapsed time is throughput;
elapsed time divided by request count is an aggregate mean cost. A median over
run-level costs is not the p50 of individual-window latencies. The reported
approximately 1.15~ms at $N=50$ is not a physical, cohort-independent
segment-decode floor.

In our profiling, sorting by sample removes the observed nonzero median of
metadata-batch reloads in the tested configurations. A zero median does not mean
reloads never occurred. Selected zstd-counter medians coincide after sorting, but
allocation and input-byte medians still vary. No equal sorted-order latency,
absence of all $N$ dependence, or causal timing effect follows from those
counters alone.

In our profiling, median assembled output is about 60 thousand bytes per request
across the tested cohorts, not an invariant for every query.
\texttt{alloc\_new\_bytes} denotes cumulative bytes requested at instrumented
allocation sites, not peak RSS or unique resident bytes. Profiler samples
attributed to \texttt{deserialize\_contig\_details} are not automatically the
fraction of elapsed query time caused by metadata work. Conditional timing of
requests with and without reload is needed for that causal claim.

\subsection{Variability and experimental units}

M6 publishes within-run-series ranges. At $N=558$, the AGC one-thread window
series has nine timed durations from 376.398550 to 378.681006~s, median
377.792440~s. These are repeated batch runs, not independent machines or
datasets. The reported integer throughput is rounded; it should not be inverted
to recover an exact latency.

GPU throughput uses five calls within one run per configuration; between-session
variability was not estimated by that procedure. The CPU predictor uses many
windows within a workload, which likewise does not establish between-session
variability. Small differences between resident forms are observations, not
evidence of a stable ranking. No confidence intervals are manufactured from
rounded medians.

\section{Reproducible archives and independent implementation}

\subsection{Encoder determinism is conditional}

The archive identity experiment covers two specific manifest archives and the
documented compiler, flags, thread counts and FMA-enabled architecture settings.
It does not prove independence from every architecture. For the reported q4k
archive, disabling FMA contraction changes the size from 21{,}765{,}741 to
21{,}765{,}975 bytes while preserving correct FASTA recovery. Reproduction must
pin the compiler, linked libraries, floating-point settings and full command. The
mechanism of differences under another compiler is not inferred.

\subsection{Specification package, not prose alone}

The clean-room developer received a specification, \texttt{EXPECTED.json}, test
vectors, archives, fetch tables and reference hashes. The report states no access
to the reference implementation, but source exclusion must be described as
technically enforced or organisationally instructed. Developer-visible expected
outputs are not held-out tests. Some reason wording was selected after consulting
those outputs.

The package tests report 26/26 full decodes, 1{,}300/1{,}300 fetches, 9/9
refusals at the expected stage and 372/372 vector fields, with exact reason text
matching in 4/9 cases. This supports implementation compatibility on that finite
package, not uniqueness or completeness of the prose specification, nor
correctness on every malformed archive. One implementer and a Python
implementation limit generalisation.

\section{Limitations and unresolved evidence}

The principal GPU result concerns q4k and the recorded batch shape, reference and
cohort. It does not establish performance across arbitrary genomes, GPUs or
application memory budgets. GPU block integrity differs from the CPU verified
path. Conditional incremental size is not total deployment cost. M6's
random-order workload must not be generalised to sample-grouped access.

The window predictor is empirically tested, not mathematically guaranteed for
p50. The cohort-scale counter experiment supports investigation of locality but
does not by itself prove the reported model failure or its cause.

No finished figures are claimed in this manuscript. Essential results are in text
and tables; placeholder figures from an earlier draft have been removed rather
than presented as completed evidence. Future figures must be generated from the
pinned source data without altering historical verdicts.

\section{Conclusion}

The main result is a GPU-resident representation of 558 assemblies with explicit
persistent-pool accounts and separately identified C1 and public Colab
measurements. Compact representation reduces the pool footprint; it does not
imply that an entire training application fits that footprint. CPU comparison
shows a workload-specific access/size trade-off with AGC, not universal codec
superiority.

The expected-work derivation explains why $W+Q-1$ is a useful quantity under
fixed-granule assumptions. It does not turn sequential throughput into a theorem
about random-window medians. Calibration cancels explicit $Q$ from the linear
increment, and cohort locality can change the measured cost. Independent
verification, timer definitions, complete size accounts and experiment-specific
provenance remain necessary to evaluate the claims.

\section{Data, code and audit availability}

The supplied evidence snapshot includes the cited text logs and source excerpts,
their local SHA-256 values and upstream Git blob identifiers. The repository
snapshot is pinned; a code snapshot is not necessarily the code revision used by
every experiment. Manuscript source tags resolve to the files in that snapshot.
A numerical occurrence inventory enumerates numerical occurrences, locations,
source tags, log locators and evidence status. Mathematical constants,
identifiers and editorial numbering are classified separately from measurements.
An inventory check verifies coverage, source bytes and local evidence links; it
does not certify the semantics of every number, and it is not a claim that every
experiment was independently rerun.

The source repository is panvram.\footnote{%
\url{https://github.com/yasha1971-coder/panvram}}
\texttt{FORMAT.md}, \texttt{EVIDENCE.md} and \texttt{DATASET.md} are
repository-root documents, not files under \texttt{logs/}. The $N=558$ model
verdict is published under commit
\texttt{8f8e9e1ccc328f07bc3cbfe40f0317db33e101a4},
\texttt{logs/verdict-n558-v1/}. The HPRC umbrella record is
PRJNA730823. Neither citation implies institutional endorsement. An archival
deposit of this manuscript and its evidence package is at Zenodo,
DOI~10.5281/zenodo.23449044.

The remaining outstanding item is the exact adapter timer inventory. The M6
protocol and selected raw timing logs are included; publication of a summary or
hash list alone does not supply missing raw-window observations. This is a
revised manuscript with an explicit evidence boundary, not a claim of completed
submission readiness.

\section*{Acknowledgements}

We acknowledge the Human Pangenome Reference Consortium for the source data
resource identified by the linked BioProject, and the Telomere-to-Telomere
consortium for the reference resource. Dataset acknowledgements do not imply
review or endorsement of this work. Applicable release and use terms must be
checked for the exact manifest inputs; this manuscript does not make a blanket
public-domain claim.

\emph{AI assistance.} Code for the measurement harness and evidence package,
number auditing, and drafting/editing of the text were assisted by Claude
(Anthropic), ChatGPT (OpenAI), Gemini (Google) and Meta AI. All measurements were
run on the author's hardware; every number is traceable to a published log. The
author designed the study, checked and approved all results and the final text,
and is responsible for the content.

\appendices

\section{Historical predictor evidence, not the principal cohort result}

The recovered experiment is attributed to $N=4$. The sum of the four assemblies
is 11{,}989{,}829{,}249 canonical bases; a transcribed 12{,}009{,}829{,}249 in an
earlier summary was a source typo, corrected in panvram 8f8e9e1. The canonical
numerator is therefore settled, and the $D_Q$ and predictions in the tables below
use it.

The following historical model-B table (Table~\ref{tab:histB}) is preserved
byte-for-byte. Its source is the historical summary. Error uses prediction as
denominator.

\begin{table*}[t]
\caption{Historical model-B verdicts ($N=4$), preserved verbatim. Error uses
prediction as denominator; PASS/FAIL decisions are unchanged.}
\label{tab:histB}
\centering
\scriptsize
\setlength{\tabcolsep}{4pt}
\resizebox{\textwidth}{!}{%
\begin{tabular}{@{}llrrrllll@{}}
\toprule
format & variant & $Q$ (B) & $D_Q$ (GB/s) & $c_0$ (us) & p50 1\,KiB / err & p50 8\,KiB / err & p50 64\,KiB / err & verdict \\
\midrule
refrel3 & q4k & 4{,}096 & 2.068 & 7.25 & 9.87\,us / $+1.5\%$ & 13.37\,us / $+1.3\%$ & 42.25\,us / $+3.2\%$ & PASS \\
refrel3 & q16k & 16{,}384 & 2.085 & 6.79 & 15.12\,us / $-0.1\%$ & 18.08\,us / $-2.7\%$ & 47.58\,us / $+3.2\%$ & PASS \\
BGZF & default & 64{,}399 & 0.859 & 3.17 & 78.40\,us / $-1.2\%$ & 78.98\,us / $-10.0\%$ & 157.2\,us / $+1.8\%$ & PASS \\
BGZF & matched-g 4147 & 4{,}096 & 0.610 & 4.55 & 11.72\,us / $-9.5\%$ & 24.68\,us / $-0.1\%$ & 122.6\,us / $+3.2\%$ & PASS \\
zstd seekable & lvl 3, 16\,KiB & 16{,}384 & 0.449 & $-15.66$ & 21.03\,us / $-9.2\%$ & 22.18\,us / $-43.3\%$ & 99.07\,us / $-40.7\%$ & FAIL ($c_0<0$; 8,64\,KiB $>20\%$) \\
LZ4 indexed & 4\,MiB blocks & 4{,}192{,}248 & 0.857 & $-3{,}234$ & 1{,}659\,us / $0.0\%$ & 1{,}687\,us / $+1.2\%$ & 1{,}693\,us / $-2.4\%$ & FAIL ($c_0<0$) \\
OZSEG & l1\_w64k & 65{,}536 & 0.428 & $-26.0$ & 126.7\,us / $-2.2\%$ & 135.8\,us / $-7.2\%$ & 257.3\,us / $-8.2\%$ & FAIL ($c_0<0$) \\
OZSEG & l1\_w1m & 1{,}048{,}576 & 0.385 & $-376$ & 2{,}349\,us / $0.0\%$ & 2{,}349\,us / $-0.8\%$ & 2{,}355\,us / $-6.5\%$ & FAIL ($c_0<0$) \\
OZSEG & l3\_w64k & 65{,}536 & 0.406 & $-24.4$ & 138.6\,us / $-0.8\%$ & 138.3\,us / $-12.1\%$ & 277.8\,us / $-7.0\%$ & FAIL ($c_0<0$) \\
OZSEG & l3\_w1m & 1{,}048{,}576 & 0.386 & $-386$ & 2{,}330\,us / $-0.1\%$ & 2{,}331\,us / $-0.9\%$ & 2{,}376\,us / $-5.0\%$ & FAIL ($c_0<0$) \\
AGC & t2t & -- & -- & -- & -- & -- & -- & FAILED (no canonical granule evidence) \\
\bottomrule
\end{tabular}%
}
\end{table*}

The following historical model-A/model-B diagnostic table
(Table~\ref{tab:histA}) is also preserved. The largest reported absolute OZSEG
model-B error is 12.1\%; the sign and denominator matter.

\begin{table*}[t]
\caption{Historical model-A/model-B diagnostics ($N=4$), preserved verbatim.}
\label{tab:histA}
\centering
\scriptsize
\setlength{\tabcolsep}{3.5pt}
\resizebox{\textwidth}{!}{%
\begin{tabular}{@{}llrrrrrrrrr@{}}
\toprule
format & variant & $Q$ (B) & $D_Q$ (GB/s) & $c_0$ (us) & $W$ & p50 (us) & model A (us) & err A & model B (us) & err B \\
\midrule
zstd-seekable & lvl 3, 16\,KiB & 16{,}384 & 0.449 & $-15.7$ & 1024 & 21.03 & 38.81 & $-45.8\%$ & 23.15 & $-9.2\%$ \\
zstd-seekable & lvl 3, 16\,KiB & 16{,}384 & 0.449 & $-15.7$ & 8192 & 22.18 & 54.79 & $-59.5\%$ & 39.13 & $-43.3\%$ \\
zstd-seekable & lvl 3, 16\,KiB & 16{,}384 & 0.449 & $-15.7$ & 65536 & 99.07 & 182.62 & $-45.8\%$ & 166.97 & $-40.7\%$ \\
lz4-indexed & 4\,MiB blocks & 4{,}192{,}248 & 0.857 & $-3234.5$ & 1024 & 1658.79 & 4892.81 & $-66.1\%$ & 1658.34 & $+0.0\%$ \\
lz4-indexed & 4\,MiB blocks & 4{,}192{,}248 & 0.857 & $-3234.5$ & 8192 & 1686.51 & 4901.17 & $-65.6\%$ & 1666.71 & $+1.2\%$ \\
lz4-indexed & 4\,MiB blocks & 4{,}192{,}248 & 0.857 & $-3234.5$ & 65536 & 1692.83 & 4968.08 & $-65.9\%$ & 1733.62 & $-2.4\%$ \\
ozseg & l1\_w64k & 65{,}536 & 0.428 & $-26.0$ & 1024 & 126.70 & 155.58 & $-18.6\%$ & 129.57 & $-2.2\%$ \\
ozseg & l1\_w64k & 65{,}536 & 0.428 & $-26.0$ & 8192 & 135.84 & 172.33 & $-21.2\%$ & 146.32 & $-7.2\%$ \\
ozseg & l1\_w64k & 65{,}536 & 0.428 & $-26.0$ & 65536 & 257.26 & 306.37 & $-16.0\%$ & 280.36 & $-8.2\%$ \\
ozseg & l1\_w1m & 1{,}048{,}576 & 0.385 & $-376.5$ & 1024 & 2349.27 & 2726.73 & $-13.8\%$ & 2350.26 & $-0.0\%$ \\
ozseg & l1\_w1m & 1{,}048{,}576 & 0.385 & $-376.5$ & 8192 & 2349.06 & 2745.35 & $-14.4\%$ & 2368.89 & $-0.8\%$ \\
ozseg & l1\_w1m & 1{,}048{,}576 & 0.385 & $-376.5$ & 65536 & 2354.96 & 2894.33 & $-18.6\%$ & 2517.86 & $-6.5\%$ \\
ozseg & l3\_w64k & 65{,}536 & 0.406 & $-24.4$ & 1024 & 138.58 & 164.06 & $-15.5\%$ & 139.66 & $-0.8\%$ \\
ozseg & l3\_w64k & 65{,}536 & 0.406 & $-24.4$ & 8192 & 138.31 & 181.73 & $-23.9\%$ & 157.33 & $-12.1\%$ \\
ozseg & l3\_w64k & 65{,}536 & 0.406 & $-24.4$ & 65536 & 277.80 & 323.07 & $-14.0\%$ & 298.67 & $-7.0\%$ \\
ozseg & l3\_w1m & 1{,}048{,}576 & 0.386 & $-386.4$ & 1024 & 2330.25 & 2719.95 & $-14.3\%$ & 2333.50 & $-0.1\%$ \\
ozseg & l3\_w1m & 1{,}048{,}576 & 0.386 & $-386.4$ & 8192 & 2331.17 & 2738.52 & $-14.9\%$ & 2352.07 & $-0.9\%$ \\
ozseg & l3\_w1m & 1{,}048{,}576 & 0.386 & $-386.4$ & 65536 & 2376.01 & 2887.12 & $-17.7\%$ & 2500.68 & $-5.0\%$ \\
\bottomrule
\end{tabular}%
}
\end{table*}

\end{document}